# ICT Intervention in the Containment of the Pandemic Spread of COVID-19: An Exploratory Study

Akib Zaman [1], Muhammad Nazrul Islam [1;*], Tarannum Zaki [1], and Mohammad Sajjad Hossain [1]

[1] Department of Computer Science and Engineering, Military Institute of Science and Technology, Mirpur Cantonment, Dhaka, Bangladesh

Correspondence*:
Muhammad Nazrul Islam
nazrul@cse.mist.ac.bd

**ABSTRACT**

The objective of this article is to explore the Information and Communication Technology (ICT) interventions and its strengths, weaknesses, opportunities and threats for the containment of the pandemic spread of novel Coronavirus. The research adopted a qualitative research approach, while the study data were collected through online content review and Focus Group Discussion (FGD). Starting with a preliminary set of about 1200 electronic resources or contents, 56 were selected for review study, applying an inclusion and exclusion criteria. The review study revealed ICT interventions that include websites and dashboards, mobile applications, robotics and drones, artificial intelligence (AI), data analytic, wearable and sensor technology, social media and learning tools, and interactive voice response (IVR) as well as explored their respective usages to combat the pandemic spread of COVID-19. Later, the FGD was replicated with 22 participants and explored the possible strengths, weaknesses, opportunities, and threats (SWOT) of deploying such technologies to fight against the COVID-19 pandemic. This research not only explores the exiting status of ICT interventions to fight with the COVID-19 pandemic but also provides a number of implications for the government, practitioners, doctors, policymakers and researchers for the effective utilization of the existing ICT interventions and for the future potential research and technological development to the containment of the pandemic spread of COVID-19 and future pandemics.

## 1 INTRODUCTION

The newly discovered Coronavirus caused the COVID-19 or the Coronavirus disease bearing the symptoms of fever, dry cough, tiredness, aches and pains, sore throat and shortness of breath (1). The disease was first reported in December 2019 in Wuhan, China (2). Within the end of January, over 7,800 cases were being reported from different countries including Asia, Europe, USA, Canada and many more outside China when immediately World Health Organization (WHO) declared Coronavirus outbreak as global public health emergency (3). As the cases began to grow rapidly and healthcare systems became unable to handle the condition, in the following months with 118,000 cases in more than 110 countries over the world,

WHO declared COVID-19 as pandemic (4). As of 10 April'20, COVID-19 has spread in 210 countries and territories affecting over one and half million people, and claiming more than 100 thousands deaths (5). Figure 1 shows the percentage of total infected cases of COVID-19 around the world (6).

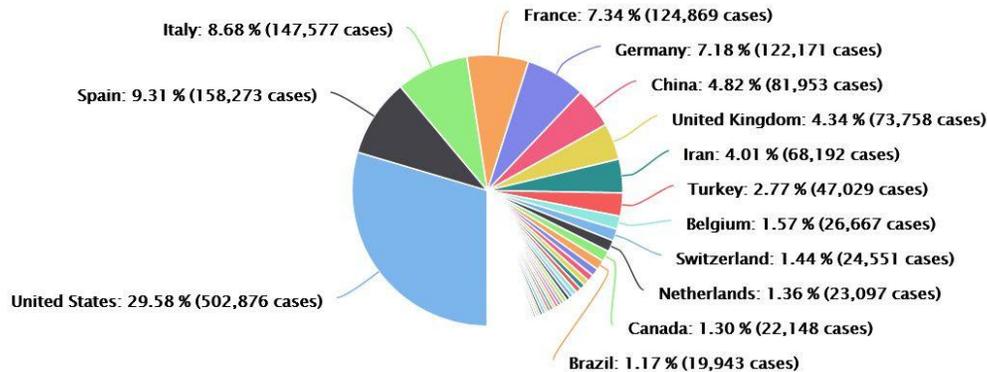

Figure 1. Percentage of infected cases around the world (6).

As clinical medicines or vaccines are under experimental processing to defend this disease, so such pandemic can be minimized with proper social responses. To engineer such social behavior, various initiatives have been taken by different health organizations to provide protective measures and awareness. WHO and CDC (Centers for Diseases Control) have notified worldwide to take necessary preparation to be safe from the exposure of this deadly disease such as practice of hygiene and social distancing, staying home, using face-mask, hand-sanitizer and disinfectants, spreading awareness among people etc. (7, 8).

Since the whole world is fighting against the pandemic spread of COVID-19, the role of Information and Communications Technology (ICT) to enhance public awareness and prevention, surveillance, diagnosis, treatment and coordinate response for COVID-19 has become more significant. Thus, the ICT interventions can be treated as one of the most effective, widely used and popular modes around the world to fight against the pandemic spread of COVID-19. A number of ICT-based initiatives has been taken around the world; for example, developing dashboard or web portal to provide the updated statistical report on Coronavirus, digital interactive maps, awareness measures and emergency calling information or hot-line numbers (9, 10, 11). WHO has gone into partnership with different social media platforms (e.g. Facebook, WhatsApp) to provide authentic information as well as health alert messaging service (12, 13). China has introduced a number of robot and drone technologies to support medical staff, thermal imaging and temperature detection software applications, smart helmets to detect potential virus carriers. India, Singapore, Canada, USA and South Korea have taken several ICT-based development program to fight against Corona pandemic (14). Few developing countries like Bangladesh has also initiated ICT interventions in the containment of COVID-19 like developing awareness groups in social media, online system for testing Coronavirus, national Corona information portal, and the likes (15, 16, 17).

However, ICT intervention for such widespread case needs definite analysis in order to explore several open issues like how much capability ICT technology holds in developing and developed countries, what opportunities lie on the usage of ICT technology in such vulnerable situation, what could be the possible threats while depending on ICT technology to mitigate the unprecedented situation caused due to pandemic spread of novel Coronavirus. Therefore, further investigation and analysis are required to explore such issues raised due to the ICT interventions for the containment of the pandemic spread of COVID-19. Thus, the objectives of this research are to investigate the ICT interventions and their roles to fight against the pandemic spread of COVID-19, and to explore the strengths, weakness, opportunities and threats (SWOT) of ICT interventions in containment of the pandemic spread of novel Coronavirus. To attain these research objectives, a qualitative research approach following the online content review and focus group discussion were conducted to provide useful insights regarding this critical global concern.

The research is organized as follows. Firstly, the related works for this research is presented in Section 2. Next, the research methodology is described in Section 3. Then, Section 4 gives an overview on the ICT interventions in COVID-19 pandemic. After that, the study results of FGD are discussed in Section 5. Finally, the main outcomes, implications, limitations and the possibility for future research are discussed in Section 6.

## 2  RELATED WORK

Though numerous number of research has been conducted focusing to the ICT interventions, but these works mainly focused to explore how the ICT interventions support in agriculture, education, women empowerment and the like (18, 19). This section briefly introduces the earlier works that were conducted focusing to the ICT interventions in the earlier pandemic and outbreak situations.

Galaz (20) in his work "Pandemic 2.0: Can Information Technology Help Save The Planet?" highlighted how the information technology can contribute in pandemic situation. In this research, Galaz (20) discussed how the information technology can facilitate to provide early warning about the epidemic, create the networks for global health governance, provide communication support in a low cost, establish network among the social organization, and the likes during pandemic. Similarly, Pletzer (21) explored the ICT services developed for reducing the Human Immunodeficiency Virus (HIV) and Acquired Immune Deficiency Syndrome (AIDS) in context of South Africa. The study included the conventional and modern ICT applications like television, video, multimedia, web portals, mobile phones, database systems, telemedicine and tele-centers. The study highlighted the effectiveness of ICTs; and found that health and HIV prevention projects should give much focus and priorities on ICT based application despite having the digital divide in the developing countries. In another work, Amicizia et al. (22) explored the significance of ICTs (mainly the Web 2.0 and mobile applications) for immunization promotion among adolescents in the developing countries, while Jeu (23) explored the role of ICT in prevention, detection and surveillance in Novel Zoonotic Disease Outbreaks; and also investigated the relationship between ICT for Development (ICT4D) and global, national, and local prevention of zoonotic diseases with pandemic potential. However, this study highlights few scopes for future technological development that may be used for the work of pandemic prevention and protection of human health.

Some other works discuss the possible use of electronic health and mobile health technologies in mitigating the impact of pandemic outbreak (24, 25). For example, Araz et al. (26) proposed a new table-top exercise tool that works through interactive simulation and visualization technique to enhance the readiness for pandemic influenza and to identify the communication gaps between responsible authorities to work during the pandemic situation.

The literature review provide a few important concern. First, only a few articles were found that explicitly focused to the ICT interventions in the earlier pandemic and outbreak situations. Second, the earlier work showed that ICT interventions has a strong significance in the containment of the earlier pandemics. Third, none of the studies found that focused to the ICT interventions in the containment of COVID-19 pandemic;

similarly, no study was found that explore the possible strengths, weaknesses, opportunities and threats for the better deployment and future development to control, manage and mitigate the COVID-19 pandemic and future pandemics/outbreaks. Therefore, this research aimed to focus on the later concern to explore the the ICT interventions and its SWOT in the containment of the pandemic spread of COVID-19 and future pandemics.

## 3 RESEARCH METHOD

From a methodological perspective, this research follows online content review (27) and focus group discussion (FGD) (28) method. Since,there are no academic literature available in the focused area and the topic is quite new and emerging, online content review approach was followed here to attain the existing ICT intervention for the containment of the pandemic spread of novel Coronavirus. The online content were analyzed through systematic coding and interpretation. Later, the outcomes of the review study were analyzed through FGD to attain the strengths, weaknesses, opportunities and threats of ICT intervention in the pandemic spread of COVID-19. In other words, an analytical assessment through FGD was performed to reveal the SWOT (strengths, weaknesses, opportunities and threats) (29) of the ICT interventions during the pandemic spread of COVID-19.

### 3.1 Online content review

The objective of online content review is to explore the existing ICT interventions around the world to fight against the COVID-19 pandemic. The Yahoo search engine and the Google search engine were used to find the related online content. The search strings used to find the available online content were "information technology and COVID-19", "ICT intervention and Coronavirus", "ICT intervention and COVID-19", "information and communication technology and digital services and Coronavirus", "information and communication technology and digital services and COVID-19", "electronic health services and mobile applications and Coronavirus", "electronic health services and mobile applications and COVID-19", "artificial intelligence and COVID-19 and Coronavirus", "big data and data science and Coronavirus and COVID-19", "robotic and wearable technology and COVID-19", and "robotic and wearable technology and Coronavirus".

The search results produced more than 1200 sites of content. The exclusion criteria include: (a) online content that discussed the same topic, especially the news articles that focus on the same digital/ICT interventions but published in multiple newspapers; (b) online contents that are not focused to our research objectives; and (c) the contents written other than English and Bengali. After applying this inclusion-exclusion process, we finally selected 56 online resources that include electronic news articles, online press releases, websites/web contents of different organizations, and the article written as a blog.

The selected articles were reviewed systematically to extract data primarily related to the technology or digital solution used in pandemic spread of COVID-19; and the purpose of using such digital technologies in COVID-19 pandemic. Finally, the extracted data were synthesized and analyzed to explore the digital interventions in the containment of pandemic spread of COVID-19.

### 3.2 Focus group discussion

The objective of focus group discussion study was to understand the possible strengths, weaknesses, opportunities and threats of ICT interventions during the COVID-19 pandemic. Due to the present vulnerable situation, two sessions of FGD were conducted online using Zoom platform. A total of 22 (8 female and 14 male) test-subjects participated in the FGD. Among them, 16 participants were Computer Science graduate, 3 participants were Business graduate and the remaining 3 participants were Medical doctors. 18 participants from Bangladesh, 2 from USA and 2 from Canada joined in the discussion sessions. The participants average age was 27 4. All the participants are quite familiar with ICT and the current vulnerable situation due to the COVID-19 pandemic. Each session lasted for around 90 -100 minutes. The participants were invited by email, Facebook messenger or telephone to participate and to fix a suitable time. Participants' written (by email) consent to ensure anonymity and confidentiality were also taken informally through email or Facebook messenger. The R & D wing of authors' institute provided the ethics approval. Also, to record the audio video of the FGD sessions, permission were taken from each participant. For each FGD session, one of the researcher acted as the moderator and other two researchers combinedly performed the role of facilitator and moderator. At the beginning, the moderator briefly presented the purpose of this study and ensured participants that the study data will be handled anonymously and their identity or photo will not be published or disclosed in any medium. Next, the moderator briefly highlighted the existing ICT interventions and the purpose of using such technologies around the world (i.e. the outcomes of the online content review). After that, the moderator opened up the floor for sharing and discussion concerning the strengths, weakness, opportunities and threats (SWOT) of ICT interventions in the containment of the pandemic spread of novel Coronavirus. The open discussion mainly focused to explore the possible strengths to deploy such technologies, identify the possible weaknesses some countries might face, open up the future opportunities for the government, practitioners or researchers for taking necessary initiatives to take a better preparation to fight the future pandemics, and understand the existing threats that are required to address for achieving maximum benefits from ICT interventions in COVID-19 pandemic. The open discussion was mainly managed by the facilitator like if any participant wanted to talk, he/she raised hand (Zoom allow hand raise service), then all others mic would be muted unless he/she finished talking and another participant looked for the option. A screenshot of the FGD session conducted through Zoom is presented in figure 2.

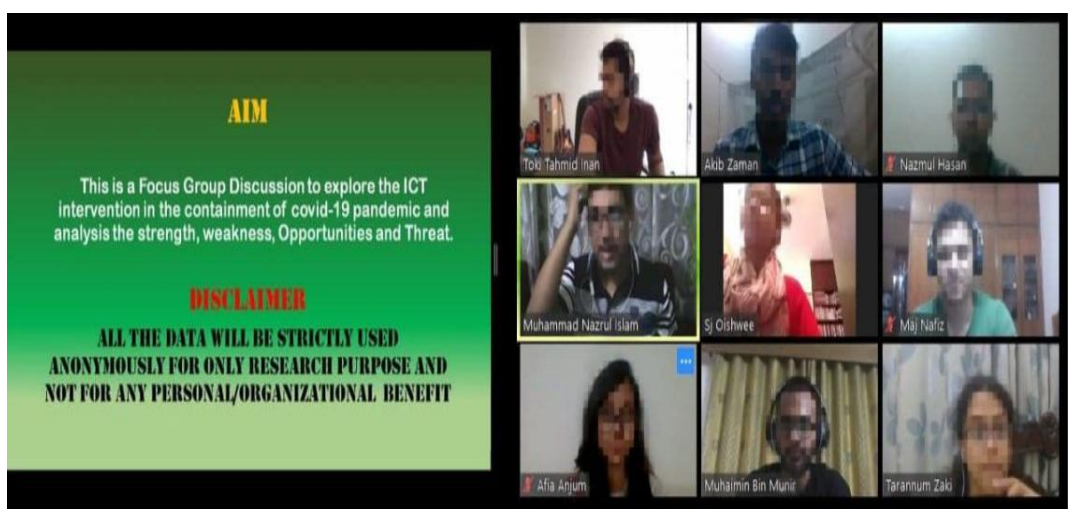

Figure 2. Screenshot of online focus group discussion session.

## 4 ICT INTERVENTIONS IN COVID-19 PANDEMIC

The review study explored the existing digital or ICT-based technologies that are being used for containment of pandemic spread of COVID-19 around the globe. Figure 3 presents the key ICT-based technologies along with three examples (for each technology). The results showed that websites and dashboards, mobile applications, robotics and drones, artificial intelligence (AI) or machine learning (ML), data analytic, wearable or sensor technology, social media and learning tools and interactive voice response (IVR) are primarily used around the world as ICT intervention to combat the pandemic spread of COVID-19 and to provide health services during this vulnerable period.

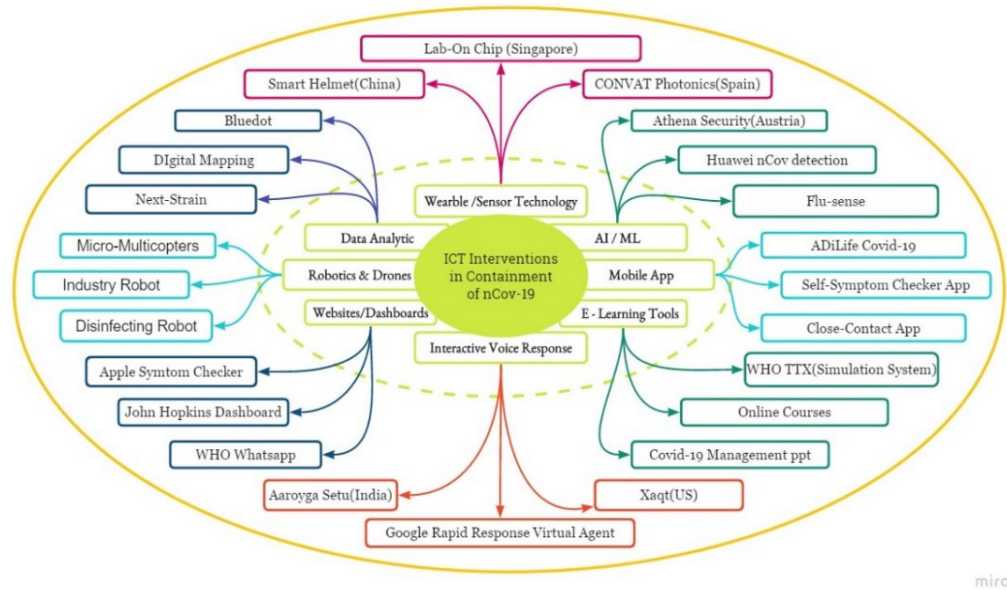

Figure 3. ICT Interventions in containment of COVID-19 pandemic.

 a) Websites and Dashboards - Most countries have created websites, dashboards, and national portals to fight within a local sphere mainly to provide updated Corona statistics, preventive and controlling information, government and medical declarations, awareness and mental health related information, emergency contact information, hospital map, and facilitate for self-assessment and symptom reporting for the risk of COVID-19. For example, WHO collaborated with WhatsApp (1) and CDC collaborated with Apple Symptom Checker (30) are striving to raise awareness and provide updated statistics of COVID-19. A dashboard developed and maintained by John Hopkins University (31) has used data visualization technique to provide latest scenario of the pandemic.

 b) Mobile Application - A number of mobile applications are being developed almost in every Corona affected country around the world. The main purpose of using mobile applications includes providing treatment information or services, remote monitoring and assistance to Corona infected patients, provide updated statistics on COVID-19, making people aware about Coronavirus, providing communication service including live video chatting and emergency calls, assisting to improve self-confidence and mental health during the pandemic situation, and providing preventative and controlling information. For example, ADiLife COVID-19 (32), HealthLynked COVID-19 Tracker (33), COVID Live Tracker - Corona Virus Pocket Guide (34) are providing remote assistance, health awareness and updated statistics about the spread of COVID-19. Again, China based apps like "Close contact" (35) and "Ali pay Health Code" (35) are using data analytic to provide color codes to every citizen which is actively helping to maintain the social distance and monitor public health.

 c) Robots and Drones - Several kind of robots and drones like disinfecting robot, agricultural robot, virtual human, micro-multicopter and industry robot are being used in several countries for multiple purposes that include transporting medical supplies, disinfection process (35), detection of symptoms (35), patient monitoring (36, 14) and in few cases providing physical medical support (35).

　d)　Artificial Intelligence - Artificial Intelligence (AI) is playing numerous roles to limit the human interaction and involvement to fight against the COVID-19 pandemic (37). Services like disease surveillance, early warnings and alerts, virtual healthcare assistance, diagnosis and prognosis, information verification over social media, controlling social distancing and measuring the lock-down, treatment and cures, processing and analyzing COVID-19 test samples, detecting and tracking individuals even when a person is wearing a face-mask and the likes are received from the AI-based technologies (38, 39). For example, AI based system Flu-sense (US) (40) is using machine learning and edge computing to predict the timing of the disease that to be launched within a small scale in university clinics. Another similar project was launched in Austria as Athena-Security (41) using thermal imaging and AI. On a similar note, Huawei has created AI based swift diagnosis tool (42) for COVID-19 patients and suspected cases.

　e)　Data Analytics- Big data, data analytic, predictive analytic or data science is used to provide data dashboard as well as to track, predict, control, respond to, and combat the pandemic spread of COVID-19 (39, 43). For example, the data analytic platform like Blue-dot (44) is used to predict the upcoming epidemic using AI and data analysis. On a similar note Next-strain (45) is an open dataset where the genome sequencing of novel corona virus was uploaded for data analysis purposes. Digital Mapping (46) was also created in few countries using data mining. Like in Bangladesh, ICT Division and ROBI Axiata Ltd has developed a data analytic based digital solutions that allows to perform various innovative data visualization exercises to generate insights and as such, the Government can assess probability of exposure in a given area to determine the next course of action (47).

　f)　Wearable Technology - Wearable computing or sensor technology like Smart Helmet (China) (35) and portable Lab-on-chip detection kit (Singapore) (14) are being used to detect COVID-19 cases in a mass crowd using thermal imaging, monitor and ensure quarantine of suspected patients, and measure/collect patient's health data remotely (48, 49). In Spain, a scientist team named CONVAT (the Catalan Institute of Nanoscience and Nanotechnology) has built photonics (50) system with ultra-sensitive laser sensor that detects Coronavirus at the earliest point of infection from a saliva or nasal swab in minutes.

　g)　Social Media and Learning Tools - Social Media platforms like Facebook, Twitter, Instagram, Telegram are playing a very effective role to raise awareness and spread preventive measures. Moreover, learning tools like simulation exercises (51) and management presentation (52) are being provided globally to raise resistance against the pandemic. Furthermore, online courses like Open-Who (53) has been introduced to provide quality knowledge all over the world.

　h)　Interactive Voice Responses - A wide use of Interactive Voice Responses (IVR) is introduced globally and locally to assist health system. Google launched Rapid Response virtual agent (54) and Xaqt's Coronavirus IVR (US) (55) as a service are conversational IVR that gives government agencies and NGOs the ability to provide relevant information to citizens in a responsive and proactive manner. Local use of IVR like Aaroyga Seba (56) in India has been also observed effective for healthcare assistance, reporting and data collection on COVID-19.

## 5   SWOT ANALYSIS

The outcomes of the focus group discussion were synthesized and analyzed to find out the strengths, weaknesses, opportunities, and threats of the ICT interventions in the containment of the pandemic spread of novel Coronavirus. Figure 4 presents the revealed strengths, weaknesses, opportunities and threats.

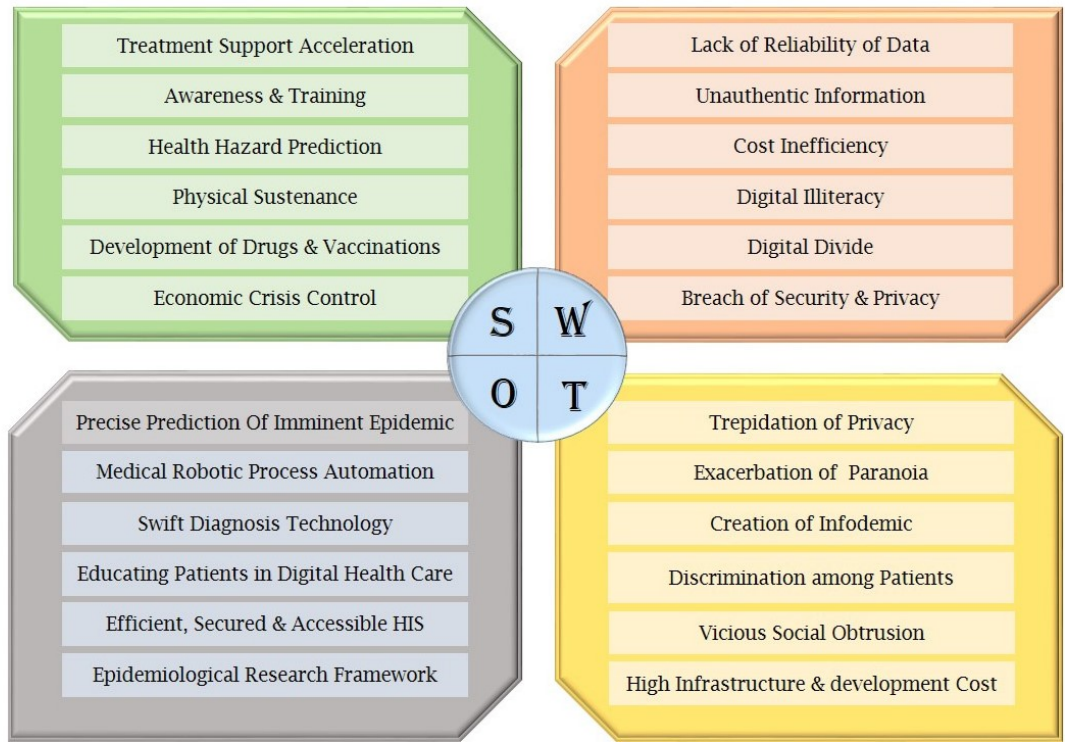

Figure 4. The SWOT of the ICT interventions in COVID-19 pandemic

## 5.1 Strengths

A total of six strengths were identified through the focus group discussion. Associations among the identified strengths and ICT interventions are shown in Figure 5. The Treatment Support Acceleration strength was cited by all the participants of FGD that includes services of treatment information circulation, remote assistance, self-assessment for COVID-19 and uncovering symptoms from a mass people through mobile applications, robotics, thermal imaging with AI, websites and social media. Another mostly cited (19 out of 22 participants) strength was Awareness & Training. Participants expressed that technologies like health websites and dashboards, online learning portals, and social media provide awareness and training on COVID-19 pandemic through data visualization of updated statistics, training of preventing measures, educating mass people about Corona, and surveillance and monitoring public health. One participant stated as "...To my understanding, it is social media that plays first role for making people aware about the pandemic of Coronavirus...". Subsequently, Health Hazard Prediction is cited by more than 80% participants as one of the important strengths of ICT intervention that includes prediction of imminent epidemic with a view to contain outbreak, and digital mapping of confirmed and suspected cases using AI and data analytic. Besides, the virtual services like providing meticulous healthcare, significant assistance in transportation of medical and humanitarian supplies, and running disinfection process with the help of modern robotic technology and drones; that have been expressed by the 16 participants (out of 22) and categorized into the strength of Physical Sustenance. Furthermore, more than one third of the participants opined that Development of Drugs and Vaccines of the virus can be achieved using positive detection of confirmed and suspected cases with the help of edge computing and AI, and Genome sequencing and epidemiological analysis using collaborative open source data collected from all over the world. Lastly, around 50% participants (10 out of 22 participants) stated about the Economic Crises Control that could be achieved using prediction of economic crisis, visualizing economic effect and recommend recovery methodologies with data analytic tool and websites. In this vein, one participant opined that "...without taking support of ICT, bringing back the economical sustainability after the pandemic period would not be possible at all...".

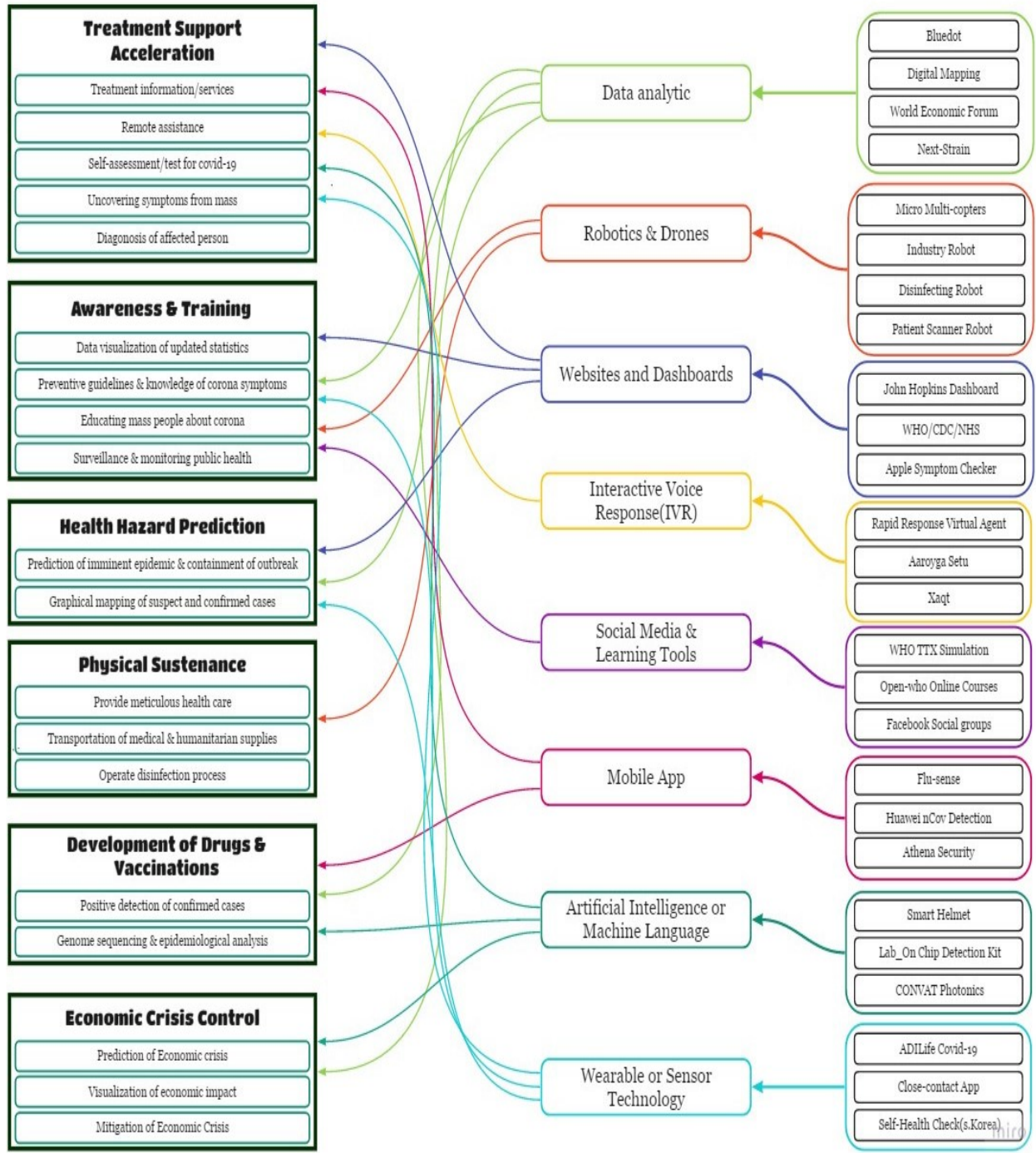

Figure 5. The strengths associated with the ICT interventions

## 5.2 Weakness

Despite having various strengths, the analysis the FGD data revealed six weaknesses of ICT interventions as presented in Figure 6.

Firstly, Lack of Reliability of data has been cited by all the participants of FGD as a prime weaknesses that can be observed mostly in the ICT domains like websites & dashboards and social media. All of the participants expressed concern that attaining data reliability is an obligation to stop infodemic during the COVID-19 pandemic. One participant stated that "...mostly in social media there are various groups who are propagating unreliable information leading into confusion and thus cater the issue of reliability...". Besides, Authenticity of Data was also cited by all the participants of FGD as a weakness in social media, IVRs and websites. Misleading and unnecessary data are creating an authenticity vacancy which is a major loophole in prevailing services. Afterwards, Cost Inefficiency is cited one of the vital weaknesses of ICT intervention by more than 75% participants where they have emphasized ignorance of considering the affordability issue of least development or developing countries due to liability of preciseness and quick manufacturing while developing robots, data analytic tools and wearable or sensor technology. Furthermore, 15 participants (out of 22) have stated Digital Divide as a weakness present in most of the domains of ICT. On this note, one of the participants stated that "...total abolition of digital illiteracy is near to impossible. Thus, people who are lacking enough digital knowledge cannot be expected to get the benefit of whatever lucrative digital systems we create...". Likewise, another participant added with the last statement that "...However, it is not only digital illiteracy, rather some of the places are deprived for the internet connectivity and limited resources specially in underdeveloped and least-developed countries...". Thus, including him, 13 participants (out of 22) acknowledged that along with digital illiteracy Digital Divide is another weakness of the prevailing ICT services. Lastly, around 55% participants (12 out of 22 participants) specified that in parallel with lack of authentication Breach of Security is another major weakness observed in mobile application and artificial intelligence.

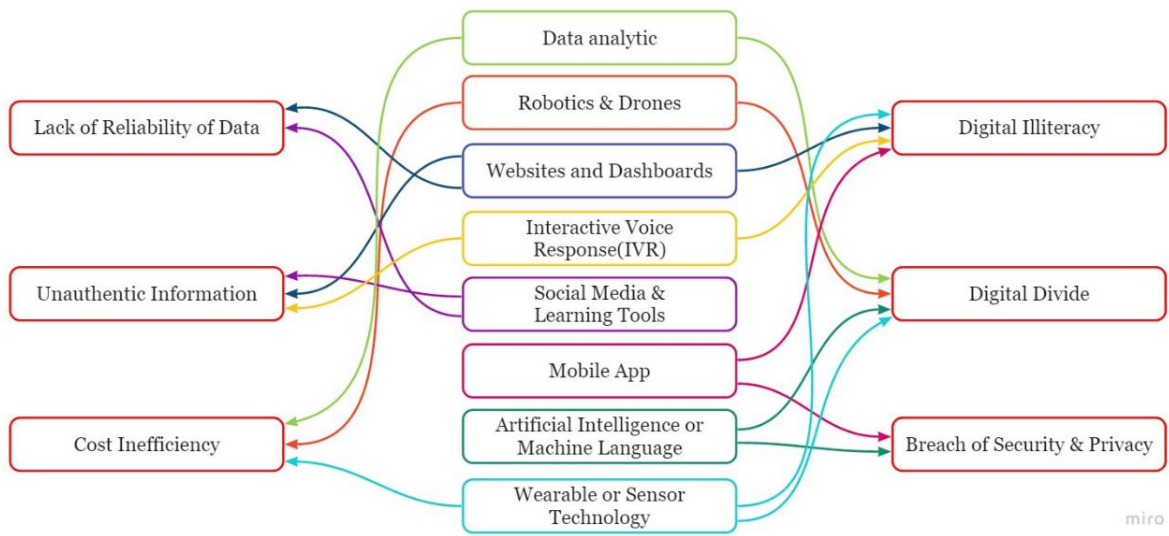

Figure 6. The weaknesses associated with the ICT interventions

## 5.3   Opportunities

Six viable opportunities to strengthen ICT intervention was found from focus group discussion. Associations among the revealed opportunities and ICT interventions are shown in Figure 7.

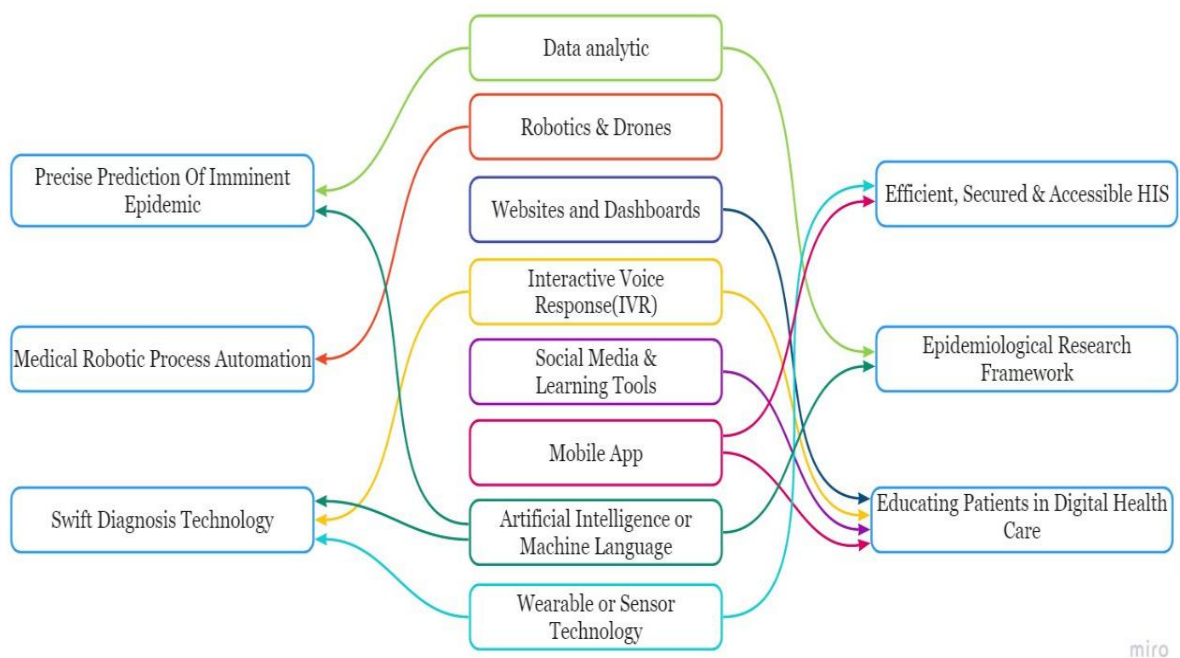

Figure 7. Association of discovered opportunities and ICT domains

All of the participants allegedly cited that Precise Prediction System of Imminent Epidemic through considerable enhancement of domains like artificial intelligence and data analytic can be the prime opportunity in days to come. Participants emphasized the importance of identifying underlined reasons of bio-hazard/epidemic with effective identification of harmful sectors which may lead into pandemic like COVID-19. One of the participant stated an interesting urge that, "...what if there was not any wet wild farming in [X country] as Global surveillance system 'Y'(fictional) has analyzed that it may create pandemic basing on info like pattern of animals cultured, maintenance procedure, types of secretion from animals etc...". In addition, More than 90% participants (20 out of 22) cited Medical Robotic Process Automation (RPA) as an opportunity to look forward. Participants discussed about how RPA can emulate human execution of repetitive processes, saving time and costs and more importantly enabling healthcare professionals to treat patients and can play a very vital role in containment of pandemic like COVID-19 and others. Moreover, 80% of the participants of FGD were in the same platform articulating that most of the COVID-19 diagnosis techniques existing now is very sluggish in nature which is prolonging the phase of pandemic. Thus, Swift Diagnosis Technology has been stated as another vital opportunity to be explored in similar situation like this in future. Subsequently, Educating Patients in Digital Health Care is crucial for the individual those are not willing and skillful to use the ICT-based health services. Educating peoples about the importance of digital health services, and providing the required level of knowledge and social encouragement about use of digital health services was stated by 15 (out of 22) participants.

Interestingly, similar number of participants 15 (out of 22) expressed Efficient, Secured & Accessible Health Information System (HIS) as an opportunity. One of the participants stand as "...Personally, I think that even in the time of epidemic it is important to maintain the secured access of personal health record (PHR) so that the social system does not collapse after the recovery from the epidemic...". Though participants shared a mixed feelings about the equality of overall desirability of new innovations in health sector; they agreed that medical professionals compared to patient advocates and administrative personnel will have a lower desirability score. This suggests a more precautious attitude of this specific interest group regarding technological innovations. Lastly, Almost 60% of the participants (13 out of 22) stated Epidemiological Research Framework as an opportunity claiming that biggest lesson out of this COVID-19 pandemic is enough preparedness was not there for tackling the microbe; thus a strong collaborative research hub is the demand of the time to fight with COVID-19 and future pandemics. On this point, one of the participants stated that, "...Genetic analysis of Human specimen can be helpful to identify which type of virus/microbe tends to attack and thus help to create preventive vaccines against them. However, this can be possible with only a strong collaborative research framework...".

## 5.4 Threats

Though with the scope numerous appealing opportunities of ICT interventions there are some probable threats which has been pigeonholed into six broad variety through focus group discussion. All the participants cited that Trepidation of Privacy as a major threat. Participants discussed that while flow of personal health information is very important to control the outbreak, perceived threat of continuous online surveillance and proliferation of this data can be disastrous. Similarly, all participants mentioned High Infrastructure & Development Cost with the opinion that in some cases prevailing health facilities are quite costly which makes it extremely difficult for the normal mass to afford the system. Thus, considering the cost-benefit ratio of the system, user may stop using them which will lead into collapse of ICT based health systems. One of the participants stated that "...it may be difficult for the least-developed and under-developed countries to afford costly technology of prevention/cure of COVID-19. Thus, it is indeed a noteworthy threat, and thus a cost effective with lower cost-benefit ratio system is always desired...". Likewise, participants discussed that with inconsistent variation in the data and huge influx of data are creating confusion; and making it very difficult for general people to crosscheck information and attain success state from the information. Thus, more than 85%, 19 (Out of 22) participants agreed that Creation of Infodemic is a threat for an effective interventions of the services. Furthermore, almost 75%, 16 out of 22 participants claimed that interference from political and religious organization to save their individual benefit is an obstruction towards free and rapid growth of ICT which has been termed as a threat named Vicious Social Obtrusion. On this vein, one of the participants stated that, "...capturing authentic data may be hazardous due to non-cooperative behavior of such organizations. It makes the control procedure of the pandemic slower and clumsy...". Interestingly, almost 60% of the participants (13 out of 22) has brought out the effect of digital divide and illiteracy, and mentioned Discrimination among Patients as a threat towards an effective interventions of ICTs. Lastly, a little less than 50% (10 out of 22) participants have talked and agreed on a threat called Exacerbation of Paranoia. One of the participants came up with a solid statement that "...While giving color codes and various segregation process, it creates a psychological impact on patients mind while being alone and imagining things before it is even happened like extreme consequences of the virus infection i.e. a hypnotic paranoia state. This is true and it happens to everybody. Perks of having too much calculation and prediction around...".

# 6 DISCUSSION AND CONCLUSIONS

This research provides an in-depth views of ICT interventions by providing (a) the type of ICT solutions that are deployed during the COVID-19 pandemic, (b) the way ICTs are being used to provide the health services, (c) the types of services or supports that are received from digital innovations, (d) explore the strengths, weakness, threats and opportunities of ICT interventions to combat with COVID-19 pandemic and to mitigate future pandemics. The outcomes of this research will greatly contribute to the practitioners, government, policy makers, doctors and individuals to aware about the ICT tools and their roles during the pandemic situations. The government of developing or infected countries may take necessary initiative to develop the affordable useful ICT-based system to provide health service and aware people to reduce the pandemic spread of COVID-19.

The outcome of this research provides implications for the potential future researches. First, investigate the critical success factors of ICT interventions during the pandemic situation in context of developing and developed countries. Second, explore the impact of ICTs on reducing and combating the COVID-19 pandemic. Third, mitigate the weaknesses and threats of ICT interventions. Finally, further research and development may focus to explore and achieve the opportunities (revealed through the SWOT analysis) to ensure the optimum benefits ICT interventions during the future pandemics.

This research has a few limitations as well. First, some related contents may be omitted due to use the specific keywords to search related content in the online review study. Second, only FGD was conducted with inadequate (22 participants) number of participants and the FGD sessions were conducted online due to the vulnerable situation. Further studies could be conducted through other methods like survey, interviews and the likes; similarly, data could be collected from a wider group of participants. Finally, it might be possible that the materials which have been used for the research may have some biases in their reporting.

However, to the best of our knowledge, no academic research or studies is conducted focusing to the ICT interventions in the COVID-19 pandemic; thus, the outcome of this research would contribute as an eye opener to the researcher, practitioners and government to take necessary further initiatives to deploy and develop such ICT or digital interventions to combat with the pandemic spread of novel Coronavirus.

## CONFLICT OF INTEREST STATEMENT

## AUTHOR CONTRIBUTIONS

## FUNDING

## ACKNOWLEDGMENTS

The authors would like to thank the participants of Focus Group Discussion whose (online) participation made the study possible despite the vulnerable situation due to the COVID-19 pandemic. Their efforts are gratefully acknowledged.

## SUPPLEMENTAL DATA

Not Applicable

## DATA AVAILABILITY STATEMENT

Not Applicable